\documentclass[twocolumn,prl,longbibliography]{revtex4-1}
\usepackage{titlesec}%
\usepackage{subfigure}%
\usepackage{amsmath}%
\usepackage{amsfonts}%
\usepackage{amssymb}%
\usepackage{graphicx}%
\usepackage{setspace}%
\usepackage{comment}%
\usepackage{dsfont}%
\usepackage{color}%
\usepackage{tikz}
\usetikzlibrary{arrows,calc}
\usepackage[pdftex,colorlinks=true,linkcolor=blue,citecolor=blue]{hyperref}

\newcommand{\ii}{\mathrm{i}}
\newcommand{\sys}{\mathcal{S}}%
\newcommand{\res}{\mathcal{R}}%

\newcommand{\ket}[1]{\left| #1 \right>} 

\titleformat{\paragraph}
{\normalfont\normalsize\bfseries}{\theparagraph}{1em}{}
\titlespacing*{\paragraph}
{0pt}{3.25ex plus 1ex minus .2ex}{1.5ex plus .2ex}

\begin{document}

\title{Bound states in the continuum of higher-order topological insulators}

\author{Wladimir A. Benalcazar}
\affiliation{Department of Physics, The Pennsylvania State University, University Park, Pennsylvania 16802, USA}
\author{Alexander Cerjan}
\affiliation{Department of Physics, The Pennsylvania State University, University Park, Pennsylvania 16802, USA}
\date{\today}

\begin{abstract}
We show that lattices with higher-order topology can support corner-localized bound states in the continuum (BICs). We propose a method for the direct identification of BICs in condensed matter settings and use it to demonstrate the existence of BICs in a concrete lattice model. Although the onset for these states is given by corner-induced filling anomalies in certain topological crystalline phases, additional symmetries are required to protect the BICs from hybridizing with their degenerate bulk states.
We demonstrate the protection mechanism for BICs in this model and show how breaking this mechanism transforms the BICs into higher-order topological resonances. Our work shows that topological states arising from the bulk-boundary correspondence in topological phases are more robust than previously expected, expanding the search space for crystalline topological phases to include those with boundary-localized BICs or resonances.
\end{abstract}

\maketitle

Topological insulators exhibit robust quantized electromagnetic phenomena with exotic boundary manifestations. A paradigmatic example is the family of topological insulators with quantized dipole moments in their bulk and charge fractionalization at their boundaries~\cite{SSH1979,Miert16,rhim17,miertOrtixFractionalCharge17}. This property of boundary charge fractionalization has recently been extended through the discovery of quantized electric multipole insulators~\cite{benalcazar2017quad,benalcazar2017quadPRB} and, more generally, $n$th-order topological insulators in $n$ dimensions, all of which host corner fractional charges in two and three dimensions (2D and 3D)~\cite{benalcazar2017quad,benalcazar2017quadPRB,Song2017,wieder2018,miertcorners,EzawaWannier19,benalcazar2019fillinganomaly,lee2019higher,sheng2019two,schindler2019}.

Among higher-order topological insulators (HOTIs) with fractional corner charges, those with additional chiral or particle-hole symmetries also host robust corner-bound states at midgap~\cite{benalcazar2014,benalcazar2017quad,Song2017}. This property makes them attractive because these topological states are easy to access experimentally due to their spectral isolation and are maximally confined~\cite{noh2018,peterson2018,ni2019}. Recently, it has also been shown that these states present nontrivial braiding properties~\cite{noh2019,pahomi2019}. 

Requiring a bulk bandgap to find spectrally isolated topological states rules out potential materials and metamaterials which otherwise possess all of the necessary crystalline symmetries to exhibit a higher-order topological phase. Yet, in principle, spectral isolation is not necessary for the existence of localized bound states. Bound states that coexist with degenerate extended ones, commonly known as bound states in the continuum (BICs), have been found across a variety of other physical systems,
including quantum systems \cite{original_1929,schult1989,moiseyev2009,cederbaum2003},
water waves \cite{ursell_trapping_1951,jones_eigenvalues_1953,callan_trapped_1991,retzler_trapped_2001,cobelli_experimental_2009,cobelli_experimental_2011},
acoustics \cite{parker_1966,parker_1967,cumpsty_1971,koch_1983,parker_excitation_1989,evans_existence_1994},
and photonics \cite{paddon_two-dimensional_2000,pacradouni_photonic_2000,ochiai_dispersion_2001,fan_analysis_2002,hsu_bloch_2013,hsu_observation_2013,yang_analytical_2014,zhen_topological_2014,zhou_perfect_2016,gao_formation_2016,kodigala_lasing_2017,zhang_extraordinary_2018,minkov_zero-index_2018,cerjan_bound_2019,plotnik_experimental_2011,weimann_compact_2013,corrielli_observation_2013,gomis-bresco_anisotropy-induced_2017,mukherjee_topological_2018}.

Thus, the natural question to consider is, do topological crystalline insulators with fractional corner charges still possess corner-localized states in the absence of a gap? And, if so, what protects these states from hybridizing with bulk states at the same energy? 
If such protected corner-localized modes do exist, they are condensed matter realizations of BICs, as they would localize to a zero-dimensional region of the system despite the existence of the background of continuum states in the bulk of the material. 

Previous studies on BICs consider systems which are coupled to scattering channels in the surrounding continuum that satisfy radiative boundary conditions, rendering their Hamiltonians non-Hermitian by allowing energy to radiate away. 
In contrast, condensed matter systems, being closed systems, have presented challenges in even defining the appropriate criteria for diagnosing the existence of BICs, which only a few previous studies have attempted to address~\cite{yang2013,ni2019,chen2019}.

In this work, we challenge the notion that boundary states of a topological phase will inevitably mix with bulk bands that are degenerate in energy with them. Instead, we show that boundary states can remain localized despite being degenerate with bulk bands and, as such, constitute condensed-matter realizations of BICs. To do so, we draw inspiration from open systems to devise a method that allows the identification of BICs in closed crystalline systems. By adding fictitious non-Hermitian terms to the Hamiltonian of the crystal, and in the correct limits, this method identifies BICs in the \emph{original} system as the isolated states with only purely real energies in the complex energy spectrum. 
Equipped with this tool, we study a concrete model of a 2D HOTI without a bulk gap at zero energy and conclusively demonstrate the existence of zero-energy corner-localized BICs. 
We further show that the protection of BICs in this lattice exists beyond separability~\cite{robnik1986,nockel1992} and depends only on the preservation of crystalline and chiral symmetries. In the absence of these symmetries (but still preserving those which protect the HOTI phase), the BICs mix with their degenerate bulk states to become higher-order topological resonances, i.e., sets of states mostly localized at corners that nevertheless also have a delocalized bulk component, and which constitute the most general spectroscopic expression of a corner-filling anomaly and corner fractional charge.

\begin{figure}[t]
\centering
\includegraphics[width=\columnwidth]{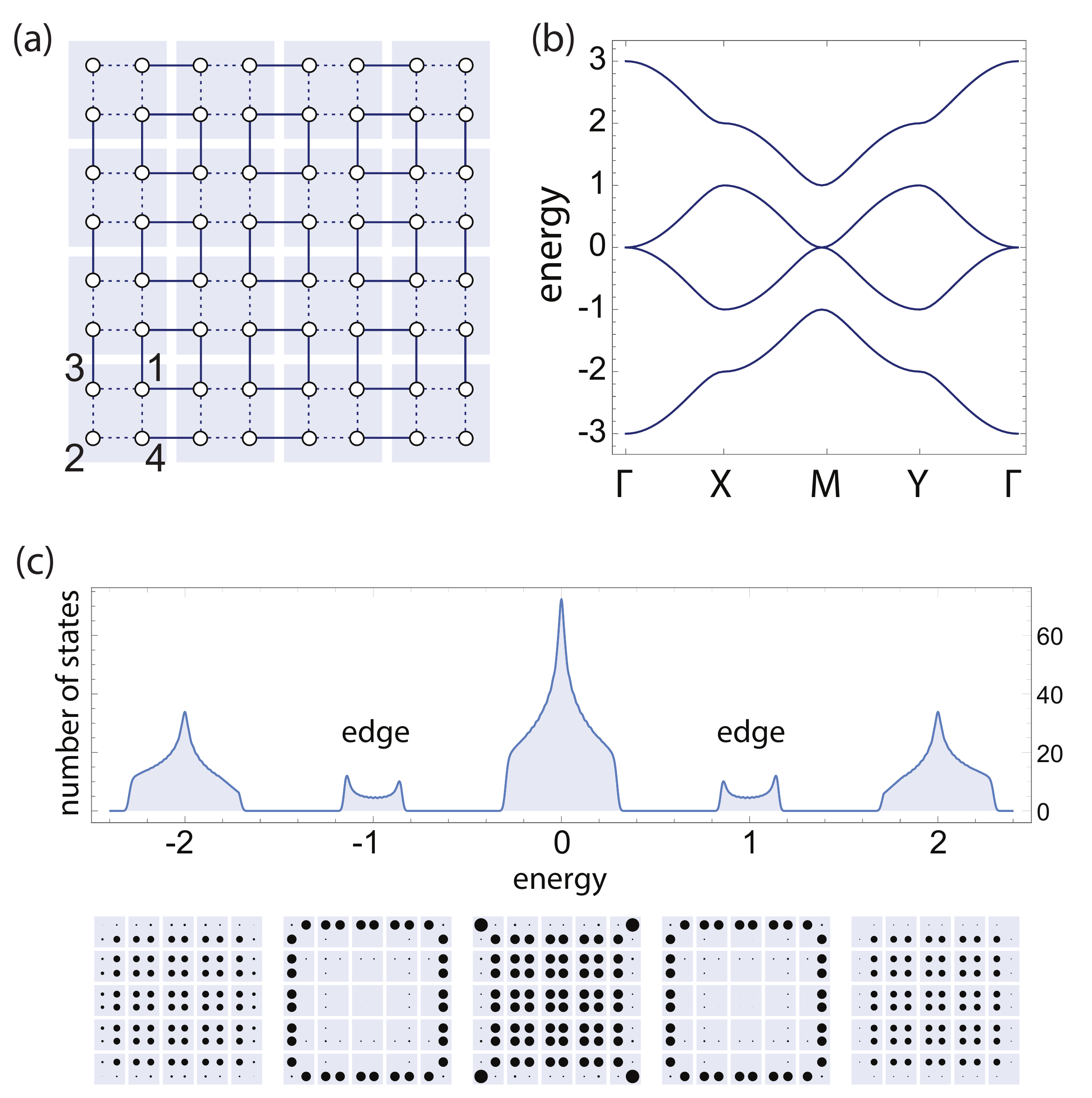}
\caption{Characterization of the model lattice with closed and open boundaries. (a) Lattice with Bloch Hamiltonian in Eq.~\ref{eq:Hamiltonian}. (b) Bulk energy bands along high-symmetry lines of the Brillouin zone for a lattice with periodic boundaries, for $t=0.5$. (c) Density of states when boundaries are open in both directions ($n=20$, $t=0.15$). The lower panels indicate the probability densities per band ($n=5$, $t=0.15$). 
}
\label{fig:lattice}
\end{figure}

\emph{Lattice and its topological phases. ---}
The lattice we consider is shown in Fig.~\ref{fig:lattice}(a) and consists of four sites per unit cell with dimerized nearest-neighbor couplings of amplitude $1$ (solid lines) and $t$ (dashed lines)~\cite{feng2017}. For the basis indicated by the numbers in Fig.~\ref{fig:lattice}(a), the Bloch Hamiltonian of the system is
\begin{align}
h({\bf k})=\left(\begin{array}{cc}
0 & Q\\
Q^\dagger & 0
\end{array}\right), \;
Q=\left(\begin{array}{cccc}
t+e^{\ii k_x} & t+e^{\ii k_y}\\
t+e^{-\ii k_y} & t+e^{-\ii k_x}
\end{array}\right).
\label{eq:Hamiltonian}
\end{align}
Its bulk energy bands are shown in Fig.~\ref{fig:lattice}(b). This Hamiltonian has chiral symmetry, $\{\Pi, h({\bf k})\}=0$, where $\Pi = \sigma_z \otimes \mathbb{I}_{2\times2}$ is the chiral operator, as well as $C_{4v}$ symmetry. As such, the spectrum is symmetric around zero energy due to chiral symmetry (see Ref.~\cite{SuppInfo}) and the two middle bands are  twofold degenerate at the ${\bf \Gamma}$ and ${\bf M}$ points of the Brillouin zone as they adopt the two-dimensional irreducible representation of $C_{4v}$. Thus, due to the simultaneous presence of chiral and $C_{4v}$ symmetries, the lattice will always have gapless bulk energy bands at zero energy.

The presence of $C_{4v}$ symmetry in the lattice distinguishes two topological phases. For $|t|<1$, all the bands in the lattice are in a topological phase, with different $C_{4v}$ ($C_{2v}$) representations at ${\bf M}$ (${\bf X}$ and ${\bf X'}$) relative to ${\bf \Gamma}$. On the other hand, for $|t|>1$, all the bands are in a trivial phase, with equal symmetry representations at all high-symmetry points (HSPs) of the Brillouin zone. The symmetry representations at all HSPs for both phases and their associated symmetry indicator topological invariants are shown in Ref.~\cite{SuppInfo}. At $|t|=1$, the phase transition occurs by closing both bulk gaps at ${\bf X}$, ${\bf X'}$ and ${\bf M}$, exchanging the representations at these three HSPs.

This model has been recently studied in the context of charge fractionalization in higher-order topological crystalline insulators~\cite{benalcazar2019fillinganomaly}. In the topological phase, the Wannier centers in all the bands localize at the maximal Wyckoff position $1b$ (corner of the unit cell), while in the trivial phase the Wannier centers in all the bands are localized at the maximal Wyckoff position $1a$ (center of the unit cell). The displacement of the Wannier centers relative to the center of their unit cells generates dipole moments per unit length quantized by $C_2$ symmetry to ${\bf P}=(\frac{e}{2},\frac{e}{2})$ in the first and fourth bands of the lattice~\cite{benalcazar2019fillinganomaly}. These quantized dipole moments are accompanied by two edge energy bands (i.e., bands with edge-localized states) spectrally isolated from the bulk energy bands [Fig.~\ref{fig:lattice}(c)]. 

In addition to the dipole moments, the topological phase has a corner-induced filling anomaly~\cite{benalcazar2019fillinganomaly} \footnote{Although a proper definition of corner-induced filling anomaly requires the vanishing of polarization in insulators, here we do not enforce this requirement as we envision the topology per band instead of the topology below a given Fermi level.} with secondary topological indices protected by $C_4$ symmetry $Q^{(4)}=\frac{1}{4}$, $\frac{1}{2}$, $\frac{1}{4}$ for the first, middle, and upper band, respectively~\cite{benalcazar2019fillinganomaly} (see Ref.~\cite{SuppInfo} for details), which capture the second-order topological character of the bands. When boundaries are open in both directions, the filling anomaly accounts for a reorganization in the number of states across bands relative to when boundaries are periodic~\cite{benalcazar2019fillinganomaly}. This reorganization is evident in the lack of homogeneity in the probability density functions shown in the lower panels of Fig.~\ref{fig:lattice}(c). In particular, the central band shows pronounced support over the corner unit cells and, as we will see, are associated with the existence of corner BICs. 

\emph{Bound states in the continuum.} In electronic systems, a nonzero filling anomaly indicates the fractionalization of the corner charge, which is the robust physical manifestation of higher-order nontrivial topology. A filling anomaly, however, does not necessarily imply the existence of zero-energy corner states. In the presence of bulk states degenerate at zero energy, as in this model, we would expect the corner and bulk states to hybridize and form corner-localized resonances, whose localization does not exponentially attenuate into the bulk completely as they would have nonzero bulk support. However, if states exponentially confined to corners exist as stand-alone eigenstates of the system despite the existence of degenerate bulk states, we will have corner-localized BICs.

\begin{figure}[t]%
\centering
\includegraphics[width=\columnwidth]{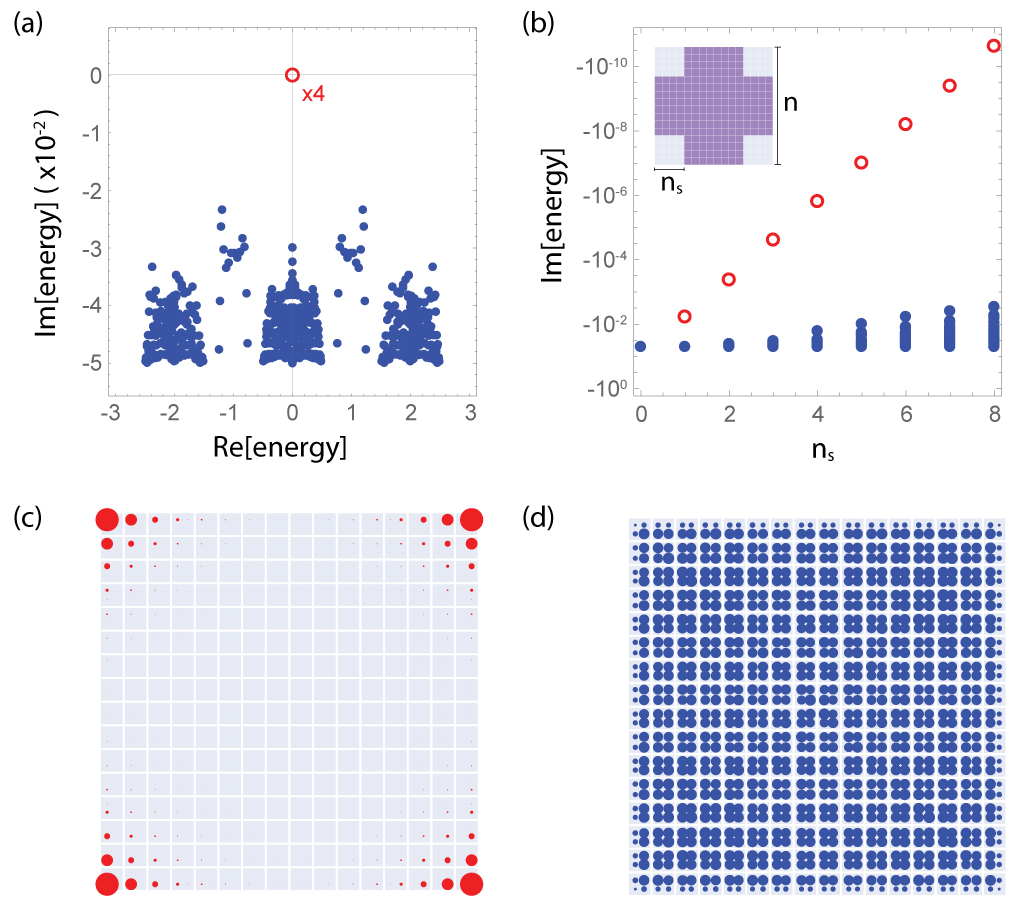}
\caption{Probing the existence of bound states in the continuum by adding the non-Hermitian term, Eq.~\ref{eq:NonHermitianReservoir}, to the lattice in Fig.~\ref{fig:lattice}(a) in the topological phase. (a) Complex energies. (b) Imaginary component of the energies as a function of system size (The inset shows the shapes of the `system' and `environment' regions in gray and purple, respectively). In (a) and (b), the red hollow circle is the four-fold degenerate energy of the bound states in the continuum with support at the corners, and the blue solid circles have eigenstates with support in bulk or edges. (c,d) Probability density function of (c) the BICs and (d) the bulk states \emph{at zero real energy}. In (c) and (d), the area of the circles is proportional to amplitude $|\psi|$ of the states. In (a), (c) and (d), $n=16$ unit cells, $n_s=3$ unit cells.  In (b) $n=32$. In all plots, $\kappa = -5\times10^{-2}$ and $t=0.25$.}
\label{fig:BIC}
\end{figure}

We can directly test for the existence of corner BICs by dividing the lattice into two regions: a small region that we leave intact which we call the `system,' $\sys$, comprised of the four square regions located at the corners of the lattice, each of size $n_s \times n_s$ unit cells, and a large region called the `environment,' $\res$, containing all of the unit cells not in $\sys$ (inset of Fig.~\ref{fig:BIC}). To the environment, we add the non-Hermitian on-site terms
\begin{align}
h_{loss}=- \ii \kappa \sum_{{\bf r} \in \mathcal{R}} \sum_{\alpha=1}^4 c^\dagger_{{\bf r},\alpha} c_{{\bf r},\alpha},  \quad 0< \kappa \ll1,
\label{eq:NonHermitianReservoir}
\end{align}
which amount to uniform losses in all the sites in the environment. If we now inject an initial wave function $\psi(0)$ into the lattice, it will evolve as $\psi(t)= e^{-\ii H t}\psi(0)$ (from now on we set $\hbar=1$), where $H$ is the Hamiltonian containing both the Hermitian Hamiltonian, Eq.~\ref{eq:Hamiltonian}, and the non-Hermitian terms, Eq.~\ref{eq:NonHermitianReservoir}.

Due to the losses in the environment and the fact that all sites in the lattice are coupled, we expect $|\psi(t)|^2$ to decrease over time. However, if corner-localized bound states exist in the continuum of the lattice, and for system sizes larger than the exponential confinement of the bound states, the losses of a wave function injected at the bound state will be heavily suppressed. This loss suppresion manifests in the propagator $e^{-\ii H t}$ by the existence of eigenstates of the Hamiltonian with close-to-real energies and bound to the corners (more precisely, the imaginary component of the complex energy of the bound states should exponentially approach zero with increasing system size).

This exact behavior of the energies of the system is observed in Fig.~\ref{fig:BIC}, which shows corner-localized bound states in the topological phase of our model. Figure~\ref{fig:BIC}(a) shows the complex energies of the Hamiltonian $H$, in which four energies are close to being purely real (red open circles), while all of the other energies have a nonvanishing imaginary component (solid blue circles). These four nearly real eigenvalues are shown in Fig.~\ref{fig:BIC}(b) to approach zero imaginary components exponentially fast with increasing system size.
As expected, the real energies have eigenstates bound to the corners [Fig.~\ref{fig:BIC}(c)]. Crucially, these corner bound states are embedded in the continuum of energies of the central bulk energy band, as can be seen in the cumulative probability density function of all eigenstates with zero real energy other than the four corner bound states, Fig.~\ref{fig:BIC}(d), which confirms that the zero-energy corner states are BICs.

\emph{BICs as a signature of the topological phase.} The penetration of the BICs into the bulk is exponentially suppressed as expected for topological states obeying a bulk-boundary correspondence\footnote{In contrast, if the BICs were the result of adding losses into the environment $\res$, we would observe their modal profiles to have a relatively uniform distribution over the entire lossless regions $\sys$}. Indeed, the corner BICs are a topological signature exclusive of the topological phase and its associated filling anomaly. When the filling anomaly vanishes, so do the BICs. 

\begin{figure}[t]%
\centering
\includegraphics[width=\columnwidth]{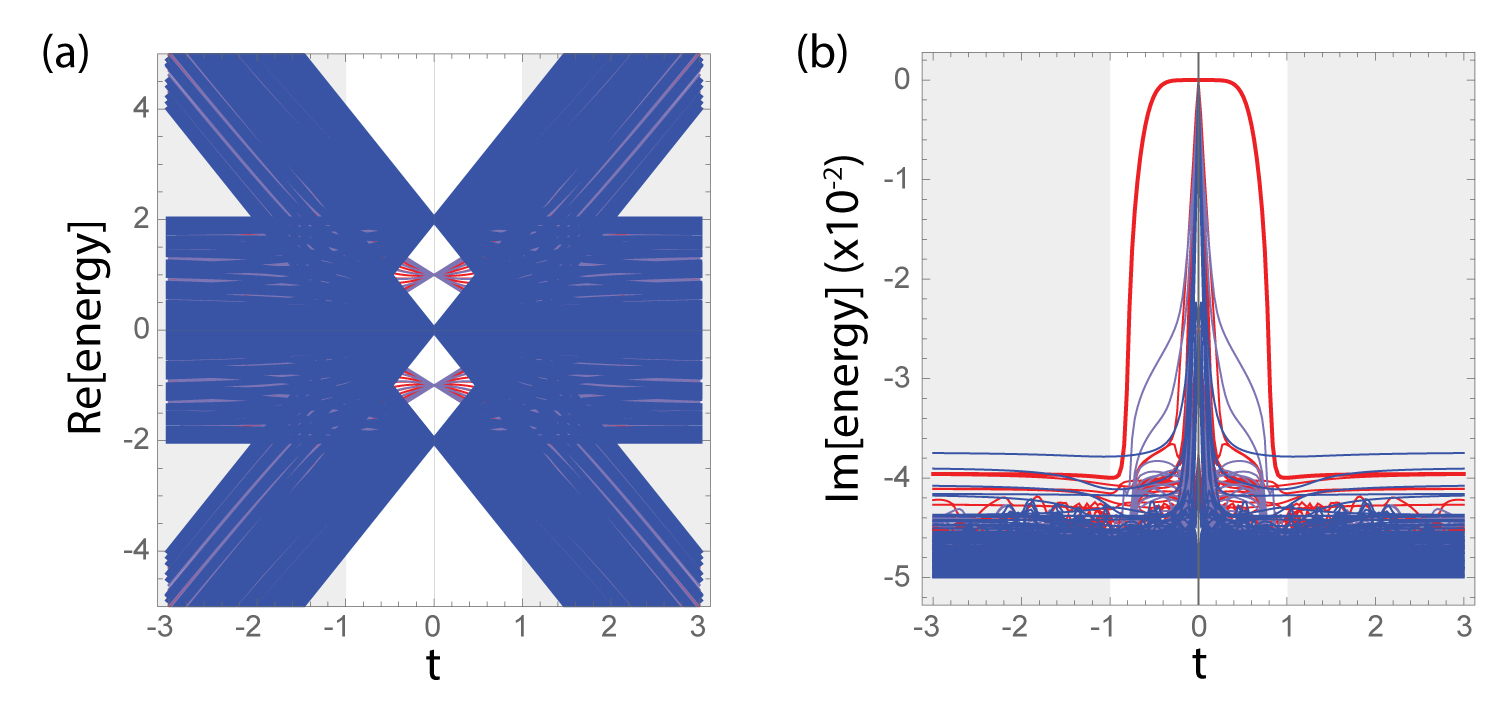}
\caption{Complex energies in the lattice as a function of the hopping amplitude $t$. Shaded and nonshaded regions correspond to the trivial  and topological phases, respectively. (a) Real component of the energies. (b) Imaginary component of the energies. Both plots show the overlapped spectra of three configurations: periodic boundaries in both directions (blue), periodic boundary only along one direction (purple), and open boundaries in both directions (red). Blue spectra are on top of purple spectra, both of which are on top of red spectra. For all plots, $n=16$, $n_s=3$, $\kappa = -5 \times 10^{-2}$.}
\label{fig:energies}
\end{figure}

This is seen in Fig.~\ref{fig:energies}, which shows the real and imaginary energies as a function of the hopping amplitude $t$. In the real spectrum [Fig.~\ref{fig:energies}(a)] it is possible to see the appearance of in-gap edge-localized states in the topological phase when boundaries are open in one direction (purple bands) \footnote{Although the plot shows that the edge states are spectrally separated from the bulk bands only for a fraction of the topological phase, they persist up to the bulk transition point $|t|=1$.} (In Fig.~\ref{fig:energies}(a), the phase transitions are not visible due to indirect gap closings in the bulk, which start to occur at $t=0.5$ [Fig.~\ref{fig:lattice}(b)]). Here, we focus on corner-bound states because they do not have any spectral isolation at any point in the real energy spectrum and their existence is far from evident. Indeed, we saw that the bound states are embedded in the continuum of the central energy band, and can be separated only in \emph{complex energy} when losses are added to the environment. Under this prescription, only the imaginary component of the spectrum allows the identification of the corner-bound states. These are shown as the red line at zero imaginary energy in the topological phase ($|t|<1$) in Fig.~\ref{fig:energies}(b). Notice the sharp transition of the BICs into lossy states as the system approaches the phase transition point ($|t|=1$). In the trivial phase ($|t|>1$), the BICs disappear as the filling anomaly vanishes. 

\emph{Symmetry protection of the BICs.} Any spatial symmetry that fixes the Wannier center of the topological phase to the maximal Wyckoff position $1b$, such as $C_2$ or $C_4$ symmetries, protects the corner filling anomaly. However, additional symmetries are required to protect the BICs from mixing with other degenerate bulk states to form resonances. In our model, both $C_{4v}$ and chiral symmetries are required to protect the BICs, as we will show now.

In the bulk, all states at zero energy take the two-dimensional representation $E$ of $C_{4v}$ (see Table~S2 in Ref.~\cite{SuppInfo} for a definition of the representations). Degenerate to these are the four corner states which, as a whole, form the representation $A_1\oplus B_2 \oplus E$. The $A_1$ and $B_2$ corner states cannot mix with the $E$ bulk states as they have incompatible symmetry representations. However, the $E$ corner and $E$ bulk states can in principle mix. Consider the combinations of corner states $\ket{C_+}=\frac{1}{2}(1,-1,\ii,-\ii)^T$ and $\ket{C_-}=\frac{1}{2}(1,-1,-\ii,\ii)^T$ that form a basis for the $E$ irreducible representation of corner states, where the entries correspond to the corner states localized at the top right, bottom left, top left, and bottom right corners, respectively. Since $\ket{C_\pm}$ form a basis for a 2D irrep, they are degenerate in energy as long as $C_{4v}$ is preserved. This basis is convenient because, in the presence of chiral symmetry, $\ket{C_\pm}$ are chiral partners of each other, i.e., $\ket{C_+}=\Pi \ket{C_-}$ and vice versa, from which it follows that these two states should have energies of opposite sign, $\epsilon, -\epsilon$ (see Ref.~\cite{SuppInfo}). Thus, under $C_{4v}$ and chiral symmetry, $\ket{C_\pm}$ must both have $\epsilon=0$. By the same argument, all bulk states that fall into the $E$ representation of $C_{4v}$ must have $\epsilon=0$ under chiral symmetry.

Now, consider a possible hybridization of the corner states $\ket{C_\pm}$ and the bulk states $\ket{B_\pm}$ (that form the $E$ representations of $C_{4v}$) into $\ket{\psi_1}= \alpha \left( \ket{B_+}+\beta \ket{C_\pm} \right)$, where $\alpha=1/\sqrt{1+|\beta|^2}$. Due to $C_{4v}$, there is another state $\ket{\psi_2}= \alpha \left( \ket{B_-}+\beta \ket{C_\mp} \right)$ degenerate to $\ket{\psi_1}$. The crucial observation is that $\ket{\psi_1}$ and $\ket{\psi_2}$ are chiral partners of each other, and as such these hybridized states have zero energy. Thus, the states $\ket{\psi_{1,2}}$ are merely arbitrary choices in the highly degenerate subspace of zero energy and do not represent a physical unbreakable hybridization into resonant eigenstates. The prescription for the detection of BICs that we propose here is then sufficient to isolate the corner BICs from the rest of degenerate bulk states.

\begin{figure}[t]%
\centering
\includegraphics[width=\columnwidth]{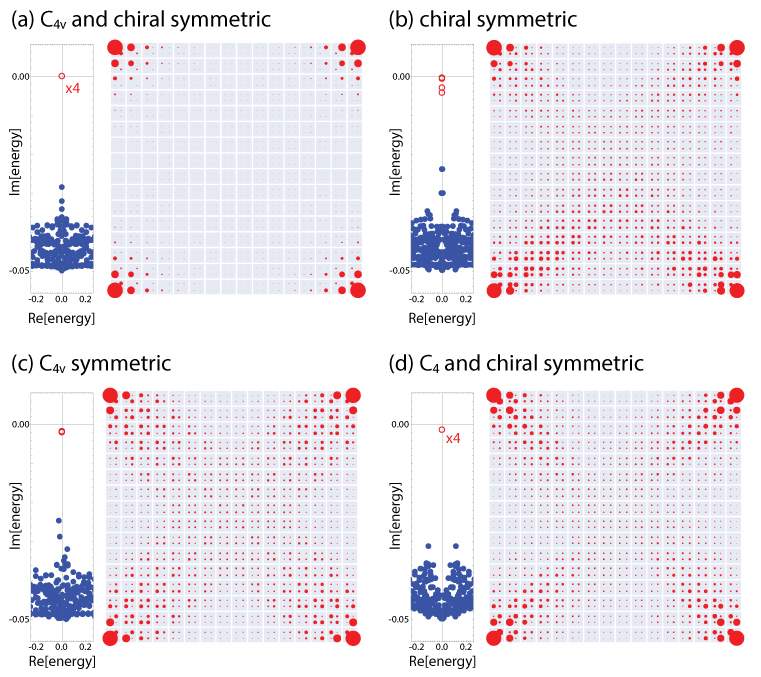}
\caption{Breaking the symmetries that protect the BICs. Energy eigenvalues (left panels) and probability densities of the four states whose energies have their imaginary components closest to zero (right panels) under perturbations that preserve certain symmetries: (a) $C_{4v}$ and chiral symmetries, (b) only chiral symmetry, (c) only $C_{4v}$ symmetry, and (d) $C_4$ and chiral symmetries. In the energy plots, the red open circles correspond to the four energies with imaginary components closest to zero (possibly degenerate). Only (a) has BICs; (b), (c), and (d) have corner-localized topological resonances. For all plots, $n=16$, $n_s=3$, $\kappa = -5 \times 10^{-2}$ and $t=0.25$.}
\label{fig:SymmetryBreaking}
\end{figure}

In the absence of either chiral or $C_{4v}$ symmetry, the hybridized states $\ket{\psi_{1,2}}$ are not pinned to zero energy and are thus free to become eigenstates of the system not susceptible of being separated into their corner and bulk constituents (Fig.~\ref{fig:SymmetryBreaking}). The inseparable hybridized states, having support in both the corner and the bulk, will eventually attenuate in the presence of loss in the environment $\res$, which manifests by a nonzero imaginary component of their energies. Some of these states are in principle long-lived as they may have more support in the corners rather than in the bulk, and thus constitute resonances of the system. In Fig.~\ref{fig:SymmetryBreaking} we show the conversion of BICs into resonances as we add perturbations to the original Hamiltonian in Eq.~\ref{eq:Hamiltonian} that break the simultaneous $C_{4v}$ and chiral symmetries down to only specific indicated symmetries. The perturbations consist of random hopping terms up to next-nearest-neighbor unit cells that nevertheless preserve the desired symmetries, as detailed in Ref.~\cite{SuppInfo}. 

In previous studies, one of the possible mechanisms for creating BICs has been attributed to the \emph{separability} of the Hamiltonian into $k_x$ and $k_y$ dependent parts, i.e., $h(k_x,k_y)=h_x(k_x)+h_y(k_y)$~\cite{robnik1986,nockel1992}. Here we show that BICs are protected beyond separability. Specifically, Fig.~\ref{fig:SymmetryBreaking}(a) has added perturbations that put the overall Hamiltonian in a nonseparable form while still hosting BICs due to the preservation of $C_{4v}$ and chiral symmetries. We also notice that in all cases in Fig.~\ref{fig:SymmetryBreaking} the filling anomaly is preserved, and in (a), (c), and (d), the Wannier centers are still fixed by symmetry to the maximal Wyckoff position $1b$. Thus, here we verify that additional symmetries to those required to protect the topological phase and its filling anomaly are required to protect BICs. Topological resonances, however, will generally exist for the symmetries that protect the topological phase, with a quality factor inversely proportional to the amplitude of the imaginary component of their energies.  The recent work of Ref.~\cite{chen2019} introduces unrestricted (i.e., symmetry-breaking) noise to their system. Thus, we expect that their numerical method for finding corner-localized states is incapable of properly distinguishing BICs from resonances.

Our demonstration that corner-localized modes exist in HOTIs even in the absence of a bulk bandgap expands the search space (design space) for topological materials (topological metamaterials). Moreover, the unique property of coexistence between BICs and bulk states offers an alternative playground for possible applications of topological phenomena.

\nocite{wannier1962,marzari1997,bradlyn2017,benalcazar2019fillinganomaly}

\begin{acknowledgments}
\emph{Acknowledgements.} We are grateful for interesting discussions with Mikael Rechtsman, Jonathan Guglielmon, S.~N. Kempkes, and Cristiane Morais Smith. W.A.B. is thankful for the support of the Eberly Postdoctoral Fellowship at the Pennsylvania State University.
A.C. acknowledges the support of the National Science Foundation Grants No. ECCS-1509546 and No. DMS-1620422 as well as the Charles E. Kaufman Foundation under Grant No. KA2017-91788.
\end{acknowledgments}
\bibliography{references}

\end{document}


\title{Supplementary Information:\\ Bound states in the continuum of higher-order topological insulators}

\author{Wladimir A. Benalcazar}
\affiliation{Department of Physics, The Pennsylvania State University, University Park, Pennsylvania 16802, USA}
\author{Alexander Cerjan}
\affiliation{Department of Physics, The Pennsylvania State University, University Park, Pennsylvania 16802, USA}

\maketitle

\beginsupplement

In this document we characterize the topological phases of the pristine lattice with Bloch Hamiltonian in Eq. 1 of the Main Text, their topological invariants, and their associated Wannier centers. We then discuss the implementation of terms that once added to the Hamiltonian in Eq. 1 of the Main Text, break certain symmetries, possibly transforming the BICs into high-order topological resonances. Finally, we present plots of the energies of the system as a function of system size to conclude that the spread of higher-order topological resonances are not due to finite size effects in the simulations. 

\subsection{Irreducible representations of the energy bands of the lattice model}
The Hamiltonian in Eq.~1 of the Main Text has $C_{4v}$ symmetry, which is generated by the simultaneous presence of $C_4$ symmetry, 
\begin{align}
\hat{r}_4 h(k_x,k_y) \hat{r}_4^\dagger =  h(k_y,-k_x),
\end{align}
and reflection symmetry,
\begin{align}
\hat{M}_x h(k_x,k_y) \hat{M}_x^\dagger =  h(-k_x,k_y),
\end{align}
both of which imply also $C_2$ symmetry as well as reflection symmetries along $y$ -denoted $M_y$- and along the two diagonals -denoted $M_{d1}$ and $M_{d2}$. The topology of the crystalline phases of this model can be diagnosed by looking at the representations that the states take at the high-symmetry points (HSPs) of the Brillouin zone. In particular, we are interested in the HSPs ${\bf \Gamma}$ and ${\bf M}$, which are invariant under the full group, $C_{4v}$, as well as the HSPs ${\bf \Gamma}$, ${\bf X}$, and  ${\bf X'}$, which are invariant only under the little group $C_{2v}$. The representations that each of these bands take at these points is given in Table~\ref{tab:BulkIrreps}.

\begin{table}[h]
\begin{tabular}{c|c|cc|ccc}
\hline
\hline
phase & bands & \multicolumn{2}{c|}{$C_{4v}$} & \multicolumn{3}{c}{$C_{2v}$}\\
& & ${\bf \Gamma}$ & ${\bf M}$ & ${\bf \Gamma}$ & ${\bf X}$ & ${\bf X'}$\\
\hline
\hline
 & 1 & $B_2$ & $A_1$ & $a_2$ & $b_1$ & $b_2$\\
$|t|<1$ & 2,3 & $E$ & $E$ & $b_1+b_2$ & $a_1 + a_2$ & $a_1+a_2$\\
& 4 & $A_1$ & $B_2$ & $a_1$ & $b_2$ & $b_1$\\
\hline
\hline
& 1 & $B_2$ & $B_2$ & $a_2$ & $a_2$ & $a_2$\\
$|t|>1$ & 2,3 & $E$ & $E$ & $b_1+b_2$ & $b_1+b_2$ & $b_1+b_2$\\
& 4 & $A_1$ & $A_1$ & $a_1$ & $a_1$ & $a_1$\\
\hline
\hline
\end{tabular}
\caption{
Symmetry representations at the high symmetry points of the BZ in both topological ($|t|<1$) and trivial ($|t|>1$) phases. Irreducible representations (irreps) at ${\bf \Gamma}$ and ${\bf M}$ are for $C_{4v}$ and irreps at ${\bf X}$ and ${\bf X'}$ are for $C_{2v}$. Irreps $A_1$, $A_2$, $B_1$, $B_2$, $a_1$, $a_2$, $b_1$, $b_2$ are one dimensional. Irrep $E$ is two-dimensional.}
\label{tab:BulkIrreps}
\end{table}

The irreducible representations in Table~\ref{tab:BulkIrreps} have the character tables detailed in Table~\ref{tab:BandRepresentations}.

\begin{table}[h]
\begin{tabular}{c|ccccc}
\hline
\hline
irrep & I & $C_2$ & $2C_4$ & $2M_v$ & $2M_d$\\
\hline
\hline
$A_1$ & 1 & 1 & 1 & 1 & 1\\
$A_2$ & 1 & 1 & 1 & -1 & -1\\
$B_1$ & 1 & 1 & -1 & 1 & -1\\
$B_2$ & 1 & 1 & -1 & -1 & 1\\
$E$ & 2 & -2 & 0 & 0 & 0\\
\hline
\end{tabular}
\quad \quad \quad
\begin{tabular}{c|ccccc}
\hline
\hline
irrep & I & $C_2$ & $M_x$ & $M_y$\\
\hline
\hline
$a_1$ & 1  & 1 & 1 & 1\\
$a_2$ & 1 & 1 & -1 & -1\\
$b_1$ & 1 & -1 & 1 & -1\\
$b_2$ & 1 & -1 & -1 & 1\\
\hline
\end{tabular}
\caption{
Character table for the $C_{4v}$ (left) and $C_{2v}$ (right) groups. The irreducible representations at the HSPs of the Brillouin zone for each energy band is shown in Table~\ref{tab:BulkIrreps}.}
\label{tab:BandRepresentations}
\end{table}
Notice that only the group $C_{4v}$ has a two-dimensional irreducible representation, $E$. This is the representation of the bulk states at zero energy and which coexist with the topological corner BICs. 

\subsection{Trivial and topological phases of the model and their Wannier centers}
\label{SuppInfo:WannierCenters}

\begin{figure}[h]%
\centering
\includegraphics[width=\columnwidth]{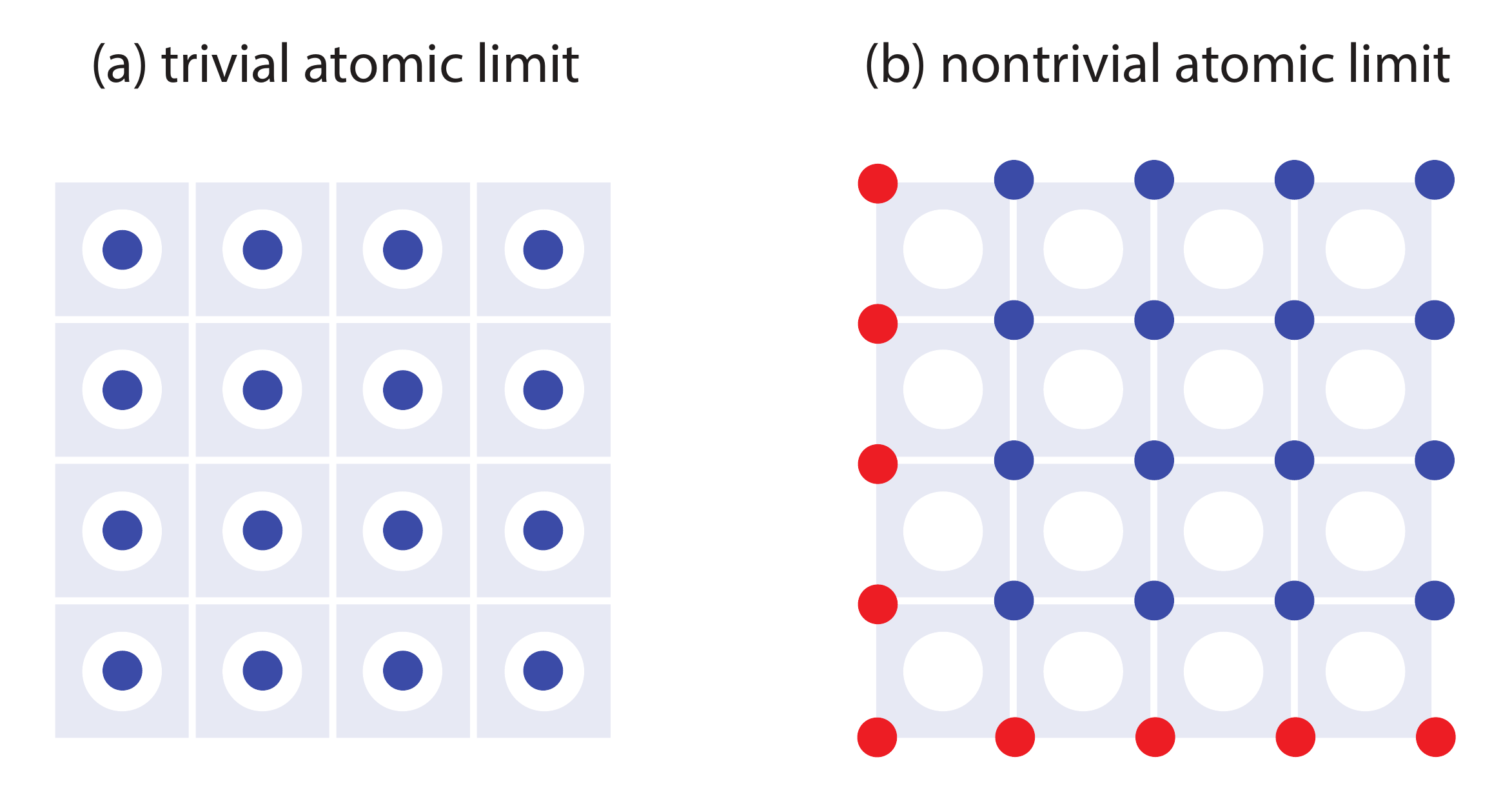}
\caption{Wannier center configuration for (a) the trivial phase, $|t|>1$ and (b) the topological phase, $|t|<1$ for all bands in the model of Eq.~1 of the Main Text. Gray squares are unit cells. Blue and red circles are the Wannier centers. White circles represent the centers of the unit cells. In (b), Wannier centers in red are in excess relative to those with closed boundaries, inducing filling anomalies that result in fractional density of states at corners.}
\label{fig:WannierCenters}
\end{figure}

In real space, the topology of the energy bands in the lattice of Fig.~1(a) in the Main Text determines the positions of their Wannier centers~\cite{wannier1962,marzari1997}. Although the Block Hamiltonian in Eq.~1 in the Main Text has $C_{4v}$ and chiral symmetries, $C_4$ or $C_2$ symmetries alone suffice to fix the positions of the Wannier centers to one of two disconnected maximal Wyckoff positions of the lattice: a `trivial' Wannier center for $|t|>1$, and a `topological' one, for $|t|<1$. These two phases are in different \emph{atomic limits}~\cite{bradlyn2017}. The trivial atomic limit ($|t|>1$) is described by Wannier centers that coincide with the centers of the unit cells [Fig.~\ref{fig:WannierCenters}(a)], and the nontrivial atomic limit ($|t|<1$) has Wannier centers at the corners of the unit cells [Fig.~\ref{fig:WannierCenters}(b)]. In particular, when boundaries are open in both directions, the nontrivial atomic limit has a mismatch in the number of Wannier centers relative to the number of unit cells [Fig.~\ref{fig:WannierCenters}(b)] which results in a fractional density of states at corners [Fig.~1(c), lower panels]. 

These two configurations can be diagnosed from symmetry indicator topological invariants, which are derived from the symmetry representations that each of the bands take at the HSPs of the Brillouin zone. Table.~\ref{tab:RotationEigenvalues} compiles the representations for $C_2$ and $C_4$ symmetries. 
\begin{table}[h]
\begin{tabular}{c|c|cc|ccc}
\hline
\hline
phase & bands & \multicolumn{2}{c|}{$C_{4}$} & \multicolumn{3}{c}{$C_{2}$}\\
& & ${\bf \Gamma}$ & ${\bf M}$ & ${\bf \Gamma}$ & ${\bf X}$ & ${\bf X'}$\\
\hline
\hline
 & 1 & $-1$ & $+1$ & $+1$ & $-1$ & $-1$\\
$|t|<1$ & 2,3 & $\pm i$ & $\pm i$ & $\{-1,-1\}$ & $\{+1,+1\}$ & $\{+1,+1\}$\\
& 4 & $+1$ & $-1$ & $+1$ & $-1$ & $-1$\\
\hline
\hline
& 1 & $+1$ & $+1$ & $+1$ & $+1$ & $+1$\\
$|t|>1$ & 2,3 & $\pm i$ & $\pm i$ & $\{-1,-1\}$ & $\{-1,-1\}$ & $\{-1,-1\}$\\
& 4 & $+1$ & $+1$ & $+1$ & $+1$ & $+1$\\
\hline
\hline
\end{tabular}
\caption{
Symmetry representations at the high symmetry points of the BZ in both topological ($|t|<1$) and trivial ($|t|>1$) phases. Irreducible representations (irreps) at ${\bf \Gamma}$ and ${\bf M}$ are for $C_{4}$ and irreps at ${\bf X}$ and ${\bf X'}$ are for $C_{2}$.}
\label{tab:RotationEigenvalues}
\end{table}
From these representations, we can calculate the symmetry indicator invariants for all the bands defined as follows,
\begin{align}
[\Pi_p]= \# \Pi_p - \#\Gamma_p,
\end{align}
where $\# \Pi_p$ is the number of states in a particular band at high symmetry point ${\bf \Pi=X}$, ${\bf Y}$, and ${\bf M}$, with eigenvalue $\Pi_p=e^{2\pi p/n}$, for $p=0\ldots n-1$, and $n=2,4$ for $C_{n=2,4}$ symmetry, respectively. 
\begin{table}[h]
\begin{tabular}{c|c|rrr|c}
\hline
\hline
phase & bands & \multicolumn{3}{c|}{Invariants} & \multicolumn{1}{c}{$Q^{(4)}_{corner}$}\\
& & ${[X_1]}$ & ${[M_1]}$ & ${[M_2]}$ & \\
\hline
\hline
 & 1 & $-1$ & $+1$ & $0$ & $\frac{1}{4}$\\
$|t|<1$ & 2,3 & $2$ & $0$ & $0$ & $\frac{1}{2}$\\
& 4 & $-1$ & $-1$ & $0$ & $\frac{1}{4}$\\
\hline
\hline
& 1 & $0$ & $0$ & $0$ & $0$\\
$|t|>1$ & 2,3 & $0$ & $0$ & $0$ & $0$\\
& 4 & $0$ & $0$ & $0$ & $0$\\
\hline
\hline
\end{tabular}
\caption{
$C_4$ symmetry indicator invariants and filling anomaly topological indices in both topological ($|t|<1$) and trivial ($|t|>1$) phases.}
\label{tab:TopologicalInvariants}
\label{tab:TopologicalInvariantsC2}
\begin{tabular}{c|c|rrr|c}
\hline
\hline
phase & bands & \multicolumn{3}{c|}{Invariants} & \multicolumn{1}{c}{$Q^{(2)}_{corner}$}\\
& & ${[X_1]}$ & ${[Y_1]}$ & ${[M_1]}$ & \\
\hline
\hline
 & 1 & $-1$ & $-1$ & $0$ & $\frac{1}{2}$\\
$|t|<1$ & 2,3 & $2$ & $2$ & $0$ & $0$\\
& 4 & $-1$ & $-1$ & $0$ & $\frac{1}{2}$\\
\hline
\hline
& 1 & $0$ & $0$ & $0$ & $0$\\
$|t|>1$ & 2,3 & $0$ & $0$ & $0$ & $0$\\
& 4 & $0$ & $0$ & $0$ & $0$\\
\hline
\hline
\end{tabular}
\caption{
$C_2$ symmetry indicator invariants and filling anomaly topological indices in both topological ($|t|<1$) and trivial ($|t|>1$) phases.}
\label{tab:TopologicalInvariantsC4}
\end{table}

From these invariants it is possible to determine the dipole moments of the bands according to the following expressions,
 \begin{align}
  {\bf P}^{(4)}=\frac{1}{2}[X_1^{(2)}]({\bf a_1}+{\bf a_2})
 \end{align}
when protected by $C_4$ symmetry and
 \begin{align}
 {\bf P}^{(2)}=\frac{1}{2}([Y_1^{(2)}]+[M_1^{(2)}]){\bf a_1}+\frac{1}{2}([X_1^{(2)}]+[M_1^{(2)}]){\bf a_2}
 \end{align}
 when protected by $C_2$ symmetry~\cite{benalcazar2019fillinganomaly}. In the topological phase,
 ${\bf P}=(\frac{e}{2},\frac{e}{2})$. With open boundaries, these moments generate an \emph{edge-induced filling anomaly}~\cite{benalcazar2019fillinganomaly}, an excess in the number of states relative to those with no boundaries. In addition, this configuration has a (nominal) corner-induced filling anomaly~\cite{benalcazar2019fillinganomaly}: an extra excess or depletion of states caused only in the presence of corners, which can be calculated via the following topological indices,
\begin{align}
Q^{(4)}_{corner} = \frac{1}{4}([X_1]+2[M_1]+3[M_2])
\end{align}
when protected by $C_4$ symmetry and 
\begin{align}
Q^{(2)}_{corner} = \frac{1}{4}(-[X_1]-[Y_1]+[M_1])
\end{align}
when protected by $C_2$ symmetry.

The filling anomaly causes existence of corner-localized states that constitute topological BICs (if additionally $C_{4v}$ and chiral symmetries are preserved) or topological resonances (if either $C_{4v}$ or chiral are broken). We emphasize that not all lattices with ${\bf P}=(\frac{e}{2},\frac{e}{2})$ have a corner-induced filling anomaly. A case in point is the lattice in Fig.~2(e) in Ref.~\onlinecite{benalcazar2019fillinganomaly}.

The symmetry indicator invariants and their topological indices for polarization and corner filling anomalies for the bands in our model are shown in Table~\ref{tab:TopologicalInvariantsC2} and \ref{tab:TopologicalInvariantsC4}. 

\subsection{Constraints on the energy spectrum due to chiral symmetry}
Consider the energy eigenstate $\ket{u}$ with energy $\epsilon$, such that
\begin{align}
h \ket{u} = \epsilon \ket{u}.
\end{align}
If the Hamiltonian $h$ has chiral symmetry, $\{h,\Pi\}=0$, then the state $\Pi \ket{u}$ is an energy eigenstate of $h$ with energy $-\epsilon$,
\begin{align}
h \Pi \ket{u} = -\Pi h \ket{u} = -\epsilon \Pi \ket{u}
\end{align}
Thus, the energies in a system with chiral symmetry come in pairs $(\epsilon,-\epsilon)$, and their states are related by the chiral operator $\Pi$. From this, it follows that states with $\epsilon=0$ are either eigenstates of $\Pi$, in which case have support only in one sublattice, or come in pairs $(\ket{u}, \Pi \ket{u})$. In the Hamiltonian of Eq.~1 in the Main Text, examples of the first case are the individual zero energy corner states, while an example of the second case is the subspace of bulk states at zero energy.

\subsection{Implementing the symmetry-breaking perturbations}
\label{sec:Perturbations}
To generate the results in Fig.~4 of the Main Text, additional hopping terms where added to the Hamiltonian of Eq.~1 in the Main Text. The overall Hamiltonian \emph{before introducing losses in the system, Eq.~2 in the Main Text}, is
\begin{align}
h_T({\bf k})=h({\bf k})+ \Delta_p h_p({\bf k}),
\end{align}
where $\Delta_p$ is the overall strength of the perturbation and
\begin{align}
h_p({\bf k})=& T_{x1} \cos k_x + T_{x2} \sin k_x + T_{y1} \cos k_y + T_{y2} \sin k_y\nonumber\\
&+T_1 \cos k_x \sin k_y + T_2 \sin k_x \cos k_y
\end{align}
is the Hamiltonian of the additional perturbation, which amounts to hopping terms up to next nearest neighbor unit cells. The $T$ matrices are all $4 \times 4$ random Hermitian matrices in which each entry has a complex value with a uniform distribution in the range $[0,1]$. In addition to obeying Hermicity, the $T$ matrices are subject to certain constraints imposed by the symmetries we are interested in preserving. In what follows we detail examples of the constraints on the $T$ matrices used for the preservation of certain symmetries:

\subsubsection{For chiral symmetry}
Under chiral symmetry, $ \Pi h({\bf k}) \Pi=-h({\bf k})$, all $T$ matrices must obey
\begin{align}
\{T,\Pi\}=0.
\end{align}

\subsubsection{For $C_4$ symmetry}
Let us first focus on the nearest neighbor $T$ matrices. To first satisfy $C_2$ symmetry, $\hat{r}_2 h(k_x,k_y)  \hat{r}_2^\dagger=  h(-k_x,-k_y)$, we require
\begin{align}
[T_{x1},\hat{r}_2]=0, \quad \{T_{x2},\hat{r}_2\}=0.
\end{align}
Now, to satisfy $C_4$ symmetry, $\hat{r}_4 h(k_x,k_y) \hat{r}_4^\dagger=  h(k_y,-k_x)$, we additionally require
\begin{align}
T_{y1} = \hat{r}_4 T_{x1} \hat{r}_4^\dagger, \quad T_{y2} = -\hat{r}_4 T_{x2} \hat{r}_4^\dagger.
\end{align}
The two next nearest neighbor $T$ matrices are odd under $C_2$ symmetry. Take first $T_1$ to obey
\begin{align}
\{T_1,\hat{r}_2\}=0,
\end{align}
and then determine $T_2$ via the constraint due to $C_4$ symmetry,
\begin{align}
T_2=-\hat{r}_4 T_2 \hat{r}_4^\dagger.
\end{align}

\subsubsection{For reflection symmetry}
Under reflection symmetry along $x$, $\hat{M}_x h(k_x,k_y) \hat{M}_x^\dagger = h(-k_x,k_y)$, four $T$ matrices are even under $M_x$ and two are odd,
\begin{align}
[T_{x1},\hat{M}_x]=0 \quad, [T_{y1},\hat{M}_x]=0, \nonumber \\ [T_{y2},\hat{M}_x]=0, \quad [T_1,\hat{M}_x]=0, \\
\{T_{x2},\hat{M}_x\}=0, \quad \{T_2,\hat{M}_x\}=0 \nonumber.
\end{align}

If more than one symmetry is to be preserved, the constraints due to each of them have to be met simultaneously. Once the $T$ matrices are chosen, an inverse Fourier transform allows to implement the hopping terms in real space. For example, the nearest neighbor perturbations along $x$ lead to
\begin{widetext}
\begin{align}
T_{x1} \cos k_x + T_{x2} \sin k_x \rightarrow \sum_{x,y} \sum_{\alpha,\beta=1}^4 c^\dagger_{(x,y),\alpha} \left( \frac{T_{x1}-\ii T_{x2}}{2}\right)_{\alpha,\beta} c_{(x+1,y),\beta}+h.c.,
\end{align}
\end{widetext}
where the sum over $x$ and $y$ run over the coordinate of unit cells in the entire lattice.

\section{Scaling of energies with lattice size}
\begin{figure}[h!]%
\centering
\includegraphics[width=\columnwidth]{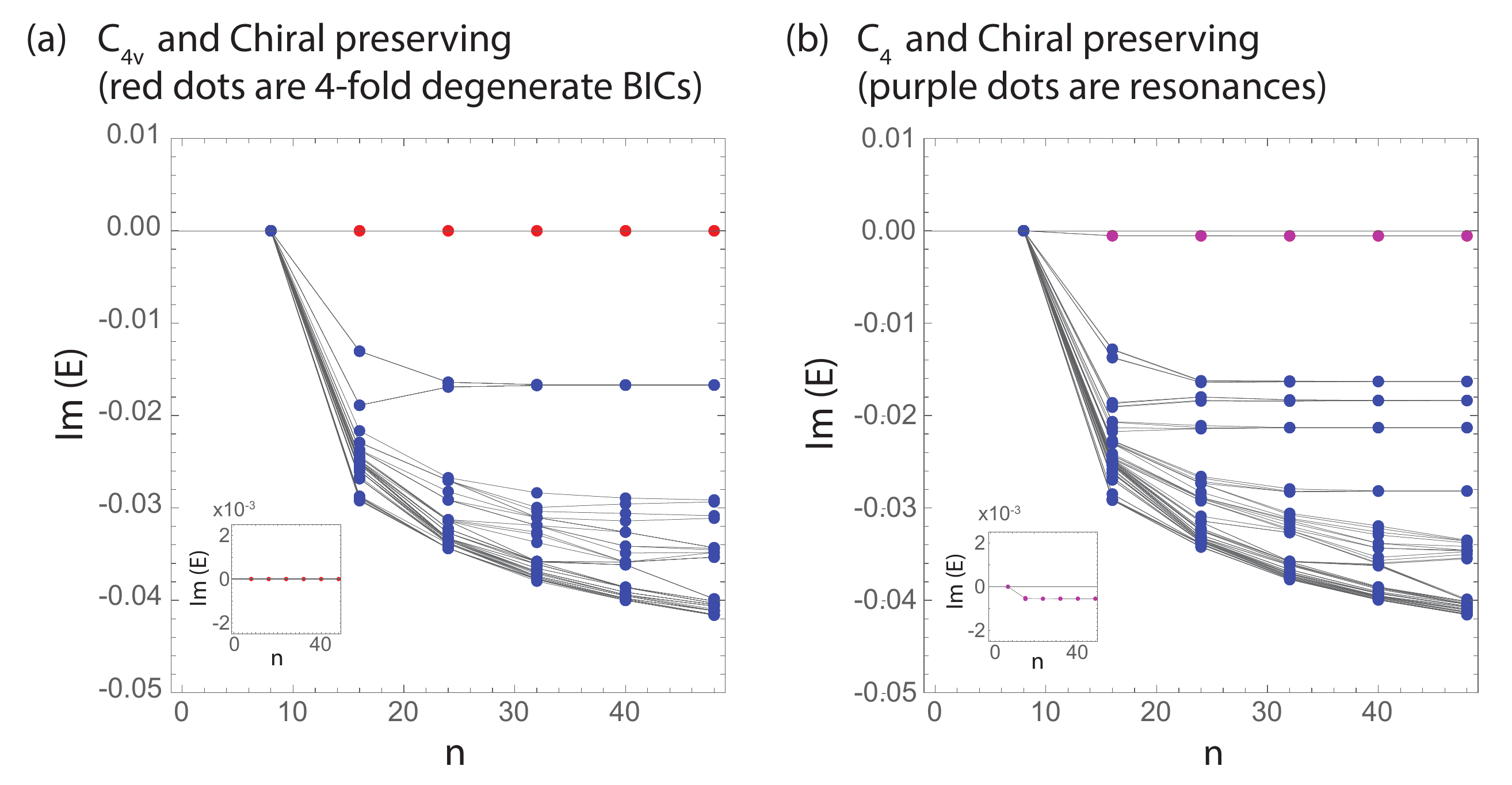}
\caption{Imaginary components of energy as a function of lattice size $n$ for a fixed value of system size $n_s=4$. (a) Lattice preserving both $C_{4v}$ and chiral symmetries. (b) A lattice that breaks $C_{4v}$ down to only $C_4$ while keeping chiral symmetry. Only in (a) there are BICs (red dots, which are 4-fold degenerate). In (b), as the lattice size increases, the imaginary components of the energies of resonances converge to a non-zero imaginary value (purple dots). The insets are zoomed in versions of (a) and (b) around Im(E)=0.}
\label{fig:scaling}
\end{figure}

Bound states exponentially penetrate into the bulk of the lattice. Resonances, on the other hand, will have both a `corner' component, which will exponentially penetrate into the lattice, and a `bulk' component, which will not. In our scheme for detection of BICs which introduces loss to the bulk but not the corners, the difference in the penetration between BICs and resonances will be manifested by zero imaginary energies for BICs and non-zero imaginary energies for resonances. To rule out the possibility that finite size effects interfere in this differentiation, in this section, we show how the values of the imaginary components of energy vary as the lattice size $n$ increases. This is shown in Fig.~\ref{fig:scaling}. In (a), we show the imaginary energies for a lattice with the original Hamiltonian, Eq.~1 of the Main Text, with added perturbations that preserve both $C_{4v}$ and chiral symmetries (implemented as described in  Section~\ref{sec:Perturbations}). Under these symmetries, corner BICs are protected and thus their imaginary energies are zero (red dots). Bulk states (blue dots), on the other hand, will acquire non-zero imaginary energies due to the losses in the bulk. In (b), perturbations are added to the original lattice, Eq.~1 of the Main Text, which break reflection symmetries. This breaks the mechanism protection of BICs, which the cease to exist as they hybridize with bulk states to form resonances. This is manifested in the fact that the imaginary energies are not zero anymore (purple dots). Notice that the values converge for lattice sizes as low as $n=16$. 

\pagebreak

\bibliography{references}